\documentclass[twocolumn, tighten, url=true]{aastex61a}

\usepackage{amsmath}
\usepackage{amssymb}
\usepackage{esint}

\usepackage[]{graphicx}

\usepackage{color}

\usepackage{natbib}

\usepackage{savesym}
\savesymbol{tablenum}
\usepackage{siunitx}
\restoresymbol{SIX}{tablenum}

\newcommand{\hyp}{\mathcal{H}}
\newcommand{\Common}{\textrm{C}}
\newcommand{\Signal}{\textrm{S}}
\newcommand{\Noise}{\textrm{N}}
\newcommand{\Hc}{\hyp^\Common}
\newcommand{\Hs}{\hyp^\Signal}
\newcommand{\Hn}{\hyp^\Noise}
\newcommand{\SignalSignal}{\Signal\Signal}
\newcommand{\Hss}{\hyp^{\SignalSignal}}
\newcommand{\Hxy}{\hyp^{XY}}
\newcommand{\Hx}{\hyp^{X}}
\newcommand{\Hy}{\hyp^{Y}}
\newcommand{\B}{\mathcal{B}}
\newcommand{\Odds}{\mathcal{O}}
\newcommand{\BOmega}{\boldsymbol{\Omega}}
\newcommand{\subA}{_{a}}
\newcommand{\subB}{_{b}}
\newcommand{\GW}{_\textrm{GW}}
\newcommand{\EM}{_\textrm{EM}}
\newcommand{\GWEM}{_\textrm{GW,EM}}
\newcommand{\Fermi}{_\textrm{\emph{Fermi}}}
\newcommand{\co}{_\textrm{c}}
\newcommand{\I}{\mathcal{I}}
\newcommand{\intDeltat}{[\Delta t]}
\newcommand{\Deltatmin}{\Delta t^{\mathrm{min}}}
\newcommand{\Deltatmax}{\Delta t^{\mathrm{max}}}
\newcommand{\ParamSpace}{\Theta}
\newcommand{\AND}{\;\textrm{and}\;}
\newcommand{\forany}{\;\textrm{for any}\;}

\begin{document}

\title{Coincident detection significance in multimessenger astronomy}
    \correspondingauthor{Gregory Ashton}

    \author{G.~Ashton}
    \email{gregory.ashton@ligo.org}
    \affiliation{Max Planck Institute for Gravitational Physics (Albert Einstein Institute), D-30167 Hannover, Germany}

    \author{E.~Burns}
    \affiliation{NASA Postdoctoral Program Fellow, Goddard Space Flight Center, Greenbelt, MD 20771, USA}

    \author{T.~\surname{Dal Canton}}
    \affiliation{NASA Postdoctoral Program Fellow, Goddard Space Flight Center, Greenbelt, MD 20771, USA}

    \author{T.~Dent}
    \affiliation{Max Planck Institute for Gravitational Physics (Albert Einstein Institute), D-30167 Hannover, Germany}

    \author{H.-B.~Eggenstein}
    \affiliation{Max Planck Institute for Gravitational Physics (Albert Einstein Institute), D-30167 Hannover, Germany}

    \author{A.~B.~Nielsen}
    \affiliation{Max Planck Institute for Gravitational Physics (Albert Einstein Institute), D-30167 Hannover, Germany}

    \author{R.~Prix}
    \affiliation{Max Planck Institute for Gravitational Physics (Albert Einstein Institute), D-30167 Hannover, Germany}

    \author{M.~Was}
    \affiliation{Laboratoire d'Annecy-le-Vieux de Physique des Particules (LAPP), Universit\'e Savoie Mont Blanc, CNRS/IN2P3, F-74941 Annecy, France}

    \author{S.~J.~Zhu}
    \affiliation{Max Planck Institute for Gravitational Physics (Albert Einstein Institute), D-14476 Potsdam-Golm, Germany}
    \affiliation{Max Planck Institute for Gravitational Physics (Albert Einstein Institute), D-30167 Hannover, Germany}

%\IfFileExists{git_tag.tex}{
%\input{git_tag.tex}
%\newcommand{\dcc}{LIGO-{\color{red}}}
%\date{\color{blue}\commitDATE; \commitIDshort-\commitSTATUS, \dcc P1700388-v1}}
%{\date{\today}}

\begin{abstract}

We derive a Bayesian criterion for assessing whether signals observed in two
separate data sets originate from a common source.  The Bayes factor for a common
vs.\ unrelated origin of signals includes an overlap integral of the posterior
distributions over the common source parameters.  Focusing on multimessenger
gravitational-wave astronomy, we apply the method to the spatial and temporal
association of independent gravitational-wave and electromagnetic (or neutrino)
observations.  As an example, we consider the coincidence between the recently
discovered gravitational-wave signal GW170817 from a binary neutron star merger
and the gamma-ray burst GRB~170817A: we find that the common source model is
enormously favored over a model describing them as unrelated signals.

\end{abstract}

\section{Introduction}

On August 17th 2017, the observation by LIGO-Virgo of GW170817, a binary
neutron star coalescence (BNS) \citep{gw170817, gwgbm}, and by \emph{Fermi} and
INTEGRAL of GRB~170817A, a short gamma-ray burst (GRB) \citep{goldstein2017,
savchenko2017}, began an unprecedented multimessenger observing campaign
\citep{multimessenger}. Detections and non-detections across the
electromagnetic (EM) spectrum and by neutrino observatories have already produced
new insights and will continue to do so for some time yet.

Many of these insights critically depend on the significance of the association
between the independent observations. Often, such significance is established
by estimating a \emph{p}-value, the probability of such an event or a more
extreme event occurring under the null hypothesis that the observations
originate from unrelated distinct sources. Specific applications include, e.g.\
\citet{gwgbm, coulter2017, soares2017} for GW170817 and its counterparts,
\citet{baret2012, aarsten2014, amon2015} for offline triggered search methods,
and \citet{urban2016monsters} for online rapid identification.  A small
\emph{p}-value demonstrates the data is inconsistent with the null hypothesis.
The \emph{p}-value cannot, though, be interpreted as the probability of the
null hypothesis itself \citep{gelman2013bayesian}. On the other hand, a large
\emph{p}-value does not necessarily imply that the null hypothesis has to be
accepted, only that it cannot be rejected \citep{gregory2005bayesian}.

We introduce a different, generic model comparison method to determine whether
two events in separate data sets are produced by a common source or by
unrelated phenomena. This Bayesian measure of significance asks fundamentally
different questions compared to the Frequentist \emph{p}-value approach: it
quantifies a \emph{degree of belief} or \emph{confidence} when comparing two
hypotheses, given a particular non-repeatable observation, while the
\emph{p}-value determines the consistency of the null hypothesis with the data
and the error rate of determining significance (which is important for initial
identification). (See \citet{finn1998} for a related discussion in the context
of detection itself).  The method is a direct comparison of the probabilities of
alternative models and does not require empirical estimates of a background
distribution for the interpretation of its result (although this may be
necessary if the assumptions about the background are not trusted).  Moreover,
the framework requires explicit statements of the necessary assumptions; in
particular, prior distributions on the relevant parameters and conditions for
which significance can be factorized for different common model parameters
(discussed later in Sec.~\ref{sec_factor}). This approach is distinct from that
of \citet{kelley2013} in which the EM data is used as prior information to
understand improvements in sensitivity for triggered searches.

In Sec.~\ref{sec_method} we introduce the method in a general context;
Eq.~\eqref{eqn_third_full} is our primary result and describes how to calculate the
Bayes factor for a common-source origin of two signals seen in separate data
streams.  In Sec.~\ref{sec_multimessenger} we focus on the application of the method to
multimessenger astronomy, considering a calculation of spatial and temporal
significance. As an example, we apply it to the gravitational-wave and
gamma-ray events GW170817 and GRB~170817A, showing that it strongly supports
the hypothesis that they originate from a common source.

\section{Generic derivation}
\label{sec_method}

\subsection{Model comparisons}

Given two detections $a$ and $b$ in different data sets $D\subA$ and $D\subB$,
we would like to assess the hypothesis that they originate from a common
source. In general, the two detections will be described by different physical
signal models $\Hs\subA$ and $\Hs\subB$, respectively. Each signal model will
imply a likelihood, a set of parameters and an associated prior for those
parameters. To assess whether they originate from a common source, the models
must share a common set of parameters $\theta\in\ParamSpace$.

We'll use notation where $\hyp(\theta)\equiv [\hyp\AND\theta]$ denotes a
hypothesis $\hyp$ with a particular choice for the parameters $\theta$, while
$\hyp$ by itself denotes a hypothesis with unknown parameters, i.e.\ ``for any
choice of parameters $\theta$''. We can formally write this as $\hyp \equiv
[\hyp(\theta) \text{ for any }\theta]$.

Then, we define the common-source hypothesis:
  \begin{align}
\Hc \equiv \left\{ \left[\Hs\subA(\theta) \AND \Hs\subB(\theta)\right]
                  \forany\theta\right\}\,.
\end{align}
We also define $\Hn_{a/ b}$  as the noise hypotheses (by which we mean any
non-signal) for each data set. Then we can define any alternative hypothesis
for which the observed detections in $a$ and $b$ are unrelated:
\begin{align}
\Hxy \equiv &\left\{
    \left[\Hx\subA(\theta\subA) \forany\theta\subA\right]
    \!\!\AND\!\!
    \left[\Hy\subB(\theta\subB) \forany \theta\subB\right]\right\}\,,
\end{align}
where $X$, $Y$ $\in\{\Noise, \Signal\}$. We write this in a general form, but
note that the noise hypothesis will not have any common model parameters. In
total, there are four possible realizations of $\Hxy$, which we consider in
detail below. However, $\Hss$ is of particular interest in this work, being two
unrelated signals from distinct sources.

These hypotheses imply priors on $\theta$ which in general differ from those
implied by $\Hs_{a/b}$ individually: if a common source can only be detected in
some subset of $\theta$, then $\Hc$ can only have prior support restricted to
this subset. If this is not true, we identify the special case
\begin{align}
P(\theta| \Hc) = P(\theta| \Hs_{a}) = P(\theta| \Hs_{b})\,.
\label{eqn_special}
\end{align}

The probability of the common-source hypothesis is given by
\begin{align}
P(\Hc| D\subA, D\subB) = \frac{P(D\subA, D\subB| \Hc)P(\Hc)}{P(D\subA, D\subB)}\,.
\label{eqn_bayes_theorem}
\end{align}
In this work, we will calculate the \emph{odds} between $\Hc$ and different
choices of $\Hxy$
\begin{align}
\begin{split}
\Odds_{\Common/XY}(D\subA, D\subB) & \equiv
\frac{P(\Hc| D\subA, D\subB)}{P(\Hxy| D\subA, D\subB)}\\
& = \B_{\Common/XY}(D\subA, D\subB)
\frac{P(\Hc)}{P(\Hxy)}\,,
\end{split}
\label{eqn_odds}
\end{align}
where
\begin{align}
\B_{\Common/XY}(D\subA, D\subB) \equiv
\frac{P(D\subA, D\subB| \Hc)}{P(D\subA, D\subB| \Hxy)}
\label{eqn_bf_definition}
\end{align}
is the \emph{Bayes factor} and $P(\Hc)/P(\Hxy)$ is the \emph{prior odds}. In
Sec.~\ref{sec_derivation} we discuss the calculation of the Bayes factor in
general. The prior odds will depend on the context, but in
Sec.~\ref{sec_final_odds} we calculate the prior odds modeling $\Hc$ and $\Hss$
as realizations of a Poisson point process.

\subsection{Derivation of the Bayes factor}
\label{sec_derivation}

If both data sets contain a signal from the same event, then they are not
independent: $P(D\subA, D\subB| \Hc) \ne P(D\subA| \Hc)P(D\subB| \Hc)$.
Instead we must compute
\begin{align}
\begin{split}
P(D\subA, D\subB| \Hc)
& =\int_\ParamSpace P(D\subA, D\subB, \theta| \Hc)\,d\theta \\
& =\int_{\ParamSpace^\Signal} P(D\subA, D\subB| \theta, \Hc)P(\theta| \Hc)\,d\theta\,,
\end{split}
\label{eqn_numerator}
\end{align}
where the domain of the integral in the second line is restricted to the prior
support of $\Hc$, namely
\begin{align}
  \label{eq:1}
  \ParamSpace^\Signal \equiv \{ \theta \in \ParamSpace \text{ where } P(\theta|\Hc)>0\}\,.
\end{align}
The need for this restriction arises because assuming that $(\theta=\theta')$
and $\Hc$ are both true would be a contradiction if $P(\theta'|\Hc)=0$, and so
$P(D|\theta',\Hc)$ would be undefined.  Rearranging the likelihood in the
integrand
\begin{align}
\begin{split}
&P(D\subA, D\subB| \theta, \Hc) \\
&= P(D\subA|D\subB, \theta, \Hc)P(D\subB| \theta, \Hc)\\
& = P(D\subA| \theta, \Hc)P(D\subB| \theta, \Hc) \\
&=
\frac{P(D\subA| \Hc)P(\theta|D\subA, \Hc)}{P(\theta| \Hc)}
\frac{P(D\subB| \Hc)P(\theta|D\subB, \Hc)}{P(\theta| \Hc)}\,,
\end{split}
\label{eqn_splitting}
\end{align}
where in the second step we have used that the likelihoods conditional on
$\theta$ can be separated for the two data sets, provided that $\theta$ is the
set of all model parameters common between the two likelihoods. In the last
step, we again used that~$P(\theta|\Hc)>0$ within the integration interval. A
subtle point is that $P(\theta| D_{a/b}, \Hc)$ is the posterior distribution
for the common model parameters (given either $D_{a/b}$) marginalized over all
other model parameters and using the prior implied by $\Hc$.

Substituting Eq.~\eqref{eqn_splitting} into Eq.~\eqref{eqn_numerator}
\begin{align}
P(D\subA, D\subB| \Hc) = P(D\subA| \Hc)P(D\subB| \Hc) \I_\theta(D\subA, D\subB)\,,
\label{eqn_dual}
\end{align}
where the \emph{posterior overlap integral}
\begin{align}
\I_\theta(D\subA, D\subB) \equiv \int_{\ParamSpace^\Signal}
\frac{P(\theta| D\subA, \Hc)P(\theta|D\subB, \Hc)}{P(\theta| \Hc)}
\,d\theta
\label{eqn_poi}
\end{align}
quantifies the agreement between the posterior distributions of $\theta$
derived independently. In this integral, the prior has the effect of setting a
scale against which the degree of overlap can be compared.

Eq.~\eqref{eqn_dual}-\eqref{eqn_poi} demonstrate how probabilities from
separate data sets combine when each provides independent inferences about a
common model parameter.

Returning to the Bayes factor, by our definition of the alternative hypothesis
\begin{align}
P(D\subA, D\subB| \Hxy)
=P(D\subA|\Hx\subA)P(D\subB| \Hy\subB)\,.
\end{align}
So from Eq.~\eqref{eqn_bf_definition},
\begin{align}
\B_{\Common/XY}(D\subA, D\subB) =
\frac{P(D\subA| \Hc)P(D\subB| \Hc)}
     {P(D\subA| \Hx\subA)P(D\subB| \Hy\subB)}
     \I_\theta(D\subA, D\subB)\,.
\label{eqn_abc}
\end{align}

We now specify three particular cases of interest for the alternative
hypothesis. First, consider $\hyp^{\Noise\Noise}$: both $a$ and $b$ are caused
by noise. Then Eq.~\eqref{eqn_abc} specializes to
\begin{align}
\B_{\Common/\Noise\Noise}(D\subA, D\subB)
&= \B_{\Common/\Noise}(D\subA)\B_{\Common/\Noise}(D\subB)
\I_\theta(D\subA, D\subB)\,.
\label{eqn_first}
\end{align}
where $\B_{\Common/\Noise}$, in analogy with Eq.~\eqref{eqn_bf_definition}, is
the common-source against noise Bayes factor. In the special case of
Eq.~\eqref{eqn_special}, it can be shown that
$\B_{\Common/\Noise}(D_{a/b})=\B_{\Signal/\Noise}(D_{a/b})$, i.e.\ the
independent signal against noise Bayes factor for each data stream.

This agrees with our intuition: if both signals are strong compared to the
background noise \emph{and} there is a good overlap of their common model
parameters (quantified by the integral), we believe they originate from a
common event.  This is a powerful result as one can compute the joint Bayes
factor from the common-source against noise Bayes factor for each detection
individually, and the posterior overlap integral of~$\theta$.
Eq.~\eqref{eqn_first} has analogous applications to the Fisher combined
probability test used in \citet{aarsten2014}.

Second, consider $\hyp^{\Signal\Noise}$: $a$ was due to a signal, but $b$ was due to
noise. For this case, Eq.~\eqref{eqn_abc} gives
\begin{align}
\B_{\Common/\Signal\Noise} =
\B_{\Common/\Noise}(D\subB) \I_\theta(D\subA, D\subB)\,.
\label{eqn_second}
\end{align}
For us to believe that detection $b$ is a real signal \emph{and} originates
from the same source as $a$, we require that the product of the Bayes factor for
common-source against noise in $b$ and the posterior overlap be large.  The
case $\B_{\Common/\Noise\Signal}$ is analogous and the same special cases apply
as mentioned previously.

Finally, consider $\Hss$, the distinct-source hypothesis: both $a$ and $b$ are of the
same nature as in the common-source hypothesis $\Hc$, but they are physically
distinct (i.e.\ they belong to unrelated sources with different parameters
$\theta\subA \not=\theta\subB$). Then,
\begin{align}
\B_{\Common/\SignalSignal}(D\subA, D\subB) \equiv
\frac{P(D\subA| \Hc)P(D\subB| \Hc)}
     {P(D\subA| \Hs\subA)P(D\subB| \Hs\subB)}
\I_\theta(D\subA, D\subB)\,.
\label{eqn_third_full}
\end{align}
This equation and the posterior overlap integral of Eq.~\eqref{eqn_poi} are the
main results of this paper. This provides a simple and intuitive way to assess
whether two detections originate from the same event, based on the
posterior overlap of their common model parameters.

In the special case of Eq.~\eqref{eqn_special}, the prefactor to the posterior
overlap integral is unity, such that
\begin{align}
\B_{\Common/\SignalSignal}(D\subA, D\subB)=
\I_\theta(D\subA, D\subB)\,.
\label{eqn_third}
\end{align}
On the other hand, when Eq.~\eqref{eqn_special} does not apply, the prefactor plays an important
role in quantifying how the restricted prior implied by $\Hc$ affects the Bayes
factor.

A similar result to Eq.~\eqref{eqn_third} was obtained independently by
\citet{haris2017} in the context of strongly lensed gravitational wave signals
from binary black hole mergers.

\subsection{Factorization of the posterior overlap integral}
\label{sec_factor}

When calculating $\I_{\theta}$, it is often convenient to factorize the posterior
overlap integral, e.g.\ $\I_{\theta}=\I_{\phi}\I_{\psi}$
where $\phi\subsetneq\theta$ and $\psi=\theta\setminus\phi$. This
factorization can only be performed, however, if
$P(\phi| \psi, D_{A/B}, \Hs) = P(\phi| D_{A/B}, \Hs)$.
There are situations in which this is the case, for example if the joint posterior
distribution is an uncorrelated multivariate normal distribution. But
generally, this will not be the case and the posterior over the full common
parameter space must be used. There are however cases where, under certain
assumptions, the integral can be approximately factorized. We will explore one
such setting in the next Section.

\section{Application to multimessenger transient astronomy}
\label{sec_multimessenger}

We now focus on the application of the above formalism to multimessenger
transient astronomy. To guide our intuition, we consider a transient
gravitational wave (GW) candidate and a detection made by an EM instrument,
although, we could just as well consider any pair of EM, GW or neutrino
detectors. Assuming that detections are made in both the GW and EM detectors
and are independently significant, we aim to calculate
$\Odds_{\Common/\SignalSignal}(D\GW, D\EM)$, the odds quantifying the
probability of the common-source hypothesis to a distinct-source hypothesis.

The Bayes factor should be calculated from all common source parameters;
typically, this will involve parameters such as a characteristic time of the
event, source direction, luminosity distance, and source orientation
\citep{troja2017, margutti2017}. Ideally, the posterior overlap integral should
be computed over the complete joint distribution of parameters since it will
not generally factorize (see Sec.~\ref{sec_factor}).

However, to illustrate the utility of the method, we will calculate the result
considering only the spatial and temporal common parameters (specifically, the
source direction $\BOmega$ and coalescence time of the BNS system $t\co$) and
make assumptions under which the posterior overlap integral may be factorized.
We also consider both observatories to be all-sky, neglecting non-isotropic and
non-stationary sensitivity. The Bayes factor can then be calculated from
Eq.~\eqref{eqn_third} since the special case of Eq.~\eqref{eqn_special}
applies.

\subsection{Example}
\label{sec_example}
To calculate the Bayes factor, Eq.~\eqref{eqn_third}, we first write down the
posterior overlap integral over the conditional joint distribution of the
spatial and temporal parameters
\begin{align}
\begin{split}
\I_{\BOmega, t\co}&(D\GW, D\EM) \\
= & \iint
\frac{P(\BOmega, t\co| D\GW, \Hs\GW)P(\BOmega, t\co| D\EM, \Hs\EM)}
{P(\BOmega, t\co| \Hs)}
\,d\BOmega dt\co\\
=&\iint
\frac{P(\BOmega| t\co, D\GW, \Hs\GW)P(\BOmega| t\co, D\EM, \Hs\EM)}
{P(\BOmega, t\co| \Hs)}\\
& \hspace{5mm} \times P(t\co| D\GW, \Hs\GW)P(t\co| D\EM, \Hs\EM)
\, d\BOmega dt\co\,.
\end{split}
\label{eqn_omegatc}
\end{align}
We will now show that this can be factorized into a spatial and temporal
overlap under the following assumptions. First, that the prior itself factors,
$P(\BOmega, t\co| \Hs) = P(\BOmega| \Hs)P(t\co| \Hs)$. Second, that $t\co$
inferred from the GW data is exactly determined, i.e.
\begin{align}
P(t\co|D\GW, \Hs\GW)=\delta(t\co - \widehat{t\co}),
\label{eqn_tco_GW}
\end{align}
where the ``hat'' indicates the observed value. Then, Eq.~\eqref{eqn_omegatc}
can be factorized as $\I_{\BOmega, t\co} = \I_{t\co}\I_{\BOmega}$, where
\begin{align}
\I_{t\co} = \frac{P(t\co=\widehat{t\co}| D\EM, \Hs\EM)}{P(t\co=\widehat{t\co}| \Hs)}
\label{eqn_temporal}
\end{align}
and
\begin{align}
\I_{\BOmega} = \int
\frac{P(\BOmega| \widehat{t\co}, D\GW, \Hs\GW)P(\BOmega| \widehat{t\co}, D\EM, \Hs\EM)}
{P(\BOmega| \Hs)} \,d\BOmega\,.
\label{eqn_spatial}
\end{align}
This factorization is exact under the two assumptions made. However, the
coalescence time is typically known with a nonzero uncertainty. For this case,
taking $\widehat{t\co}$ to be a point estimate (the mean for example), the
factorization is approximate, but applicable provided that over the uncertainty
in $t\co$, $P(t\co| D\EM, \Hs\EM)$, $P(\BOmega| t\co, D\GW, \Hs\GW)$, and
$P(\BOmega| t\co, D\EM, \Hs\EM)$ do not vary substantially. In
Sec.~\ref{sec_temporal} and Sec.~\ref{sec_spatial}, we will provide
approximations for Eq.~\eqref{eqn_temporal} and \eqref{eqn_spatial} under some
reasonable assumptions and illustrate some of the subtleties in their
calculation.

We note that a similar result to Eq.~\eqref{eqn_spatial} was previously
derived in \citet{urban2016monsters}; in particular, Eq.~(3.6) of that work is
equivalent to Eq.~\eqref{eqn_spatial} assuming an isotropic prior.  Then the
resulting joint likelihood ratio is defined using the alternative hypothesis
that went into Eq.~\eqref{eqn_second}.

\subsubsection{Temporal overlap}
\label{sec_temporal}

To evaluate Eq.~\eqref{eqn_temporal}, the temporal overlap, we first need to
consider how to compute $P(t\co| D\EM, \Hs\EM)$, the coalescence time given the
EM observations. Typically, EM observations do not directly infer $t\co$, but
some other well defined time $t\EM$, e.g.\ the time of peak
luminosity.  We therefore need to specify a model that relates these two
times. One simple model is that both signals travel at the speed of light, but
there is a delay $\Delta t=t\EM-t\co$ between the coalescence time and the EM
emission which will depend on the physics (see, e.g.\ \citet{finn1999,
abadie2012} for GRB delay time predictions), but also on how $t\EM$ is defined.
To fold these predictions into the analysis, we must specify $P(\Delta t|
\Hs)$, a prior distribution on the delay-time (at the Earth), given the model.
Assuming $\Delta t$ and $t\EM$ are independent, the posterior can be
transformed as
\begin{align}
P(t\co| D\EM, \Hs\EM) = \int
p_{t\EM}(t\co+\Delta t)P(\Delta t) d\Delta t\,,
\label{eqn_time_transform}
\end{align}
where $p_{t\EM}(t\EM)\equiv P(t\EM| D\EM, \Hs\EM)$ denotes the posterior
distribution of $t\EM$.

Having defined how to relate the time inferred by the EM data to the
coalescence time with a suitable model, we now calculate
Eq.~\eqref{eqn_temporal} under some simple assumptions. Eq.~\eqref{eqn_tco_GW}
was the first of these assumptions and was already applied in factorizing the
full posterior overlap integral. In addition, let
\begin{align}
P(t\EM| D\EM, \Hs\EM) & \equiv p_{t\EM}(t\EM) = \delta(t\EM-\widehat{t\EM})\,.
\end{align}

Next we need a prior for the delay in the GW-EM arrival time, which could be
due to differences in emission time or propagation speed of GW and EM
radiation.  For simplicity we take a uniform distribution,
\begin{align}
P(\Delta t) = \mathrm{U}^{\Deltatmax}_{\Deltatmin}(\Delta t)\,.
\end{align}
That is, the EM emission can arrive any time between a minimum and maximum
value compared to the GW-inferred coalescence time; outside of that interval,
we are certain the two events are not related. Inserting these definitions into
Eq.~\eqref{eqn_time_transform}, we obtain
\begin{align}
P(t\co| D\EM, \Hs\EM) = \mathrm{U}^{\Deltatmax}_{\Deltatmin}(\widehat{t\EM}-t\co )\,,
\label{eqn_tco_EM}
\end{align}
from which the numerator of Eq.~\eqref{eqn_temporal} can be calculated.

The prior on the coalescence time, given $\Hs$ and a co-observing time of
duration $T$, is $P(t\co| \Hs)=\mathrm{U}_0^T(t\co)$, where $t\co$ is chosen to
be zero at the start of the co-observing time.  The period $T$ should cover the
entire range of $t_c$ for which the $t\co$ posteriors (in this example,
Eq.~\eqref{eqn_tco_GW} and \eqref{eqn_tco_EM}) have nonzero support from the
data, but is otherwise an arbitrary normalization of the time prior.  Then,
Eq.~\eqref{eqn_temporal} gives
\begin{align}
\I_{t\co} = \left\{
\begin{array}{cl}
\frac{T}{\intDeltat} & \textrm{ if } (\widehat{t\co}-\widehat{t\EM}) \in [\Deltatmin, \Deltatmax]\\
0 & \textrm{ otherwise }
\end{array}
\right.\,,
\label{eqn_time_approximation_full}
\end{align}
where $\intDeltat \equiv \Deltatmax-\Deltatmin$.  The dependence on $T$ would
suggest that we can arbitrarily change the significance through the Bayes
factor by adjusting $T$. However, as will be shown in
Sec.~\ref{sec_final_odds}, this factor cancels with the prior odds
$P(\Hc)/P(\Hss)$, which depends on both $T$ and the rate of events, such that
the odds themselves are $T$-independent.  We also will see that in a particular class of
cases the temporal odds can be well approximated by this Bayes factor, setting
$T$ to the average interval between signals detectable in EM, (i.e.\ the inverse
of the rate of such signals).

\subsubsection{Spatial overlap}
\label{sec_spatial}

We now discuss calculating Eq.~\eqref{eqn_spatial}, the spatial posterior
overlap integral. The EM counterparts to GW events are expected to originate
from the same source direction and hence $\I_{\BOmega}$ can be directly
computed from Eq.~\eqref{eqn_spatial}.

\begin{figure}[b]
\centering
\includegraphics[width=0.49\textwidth]{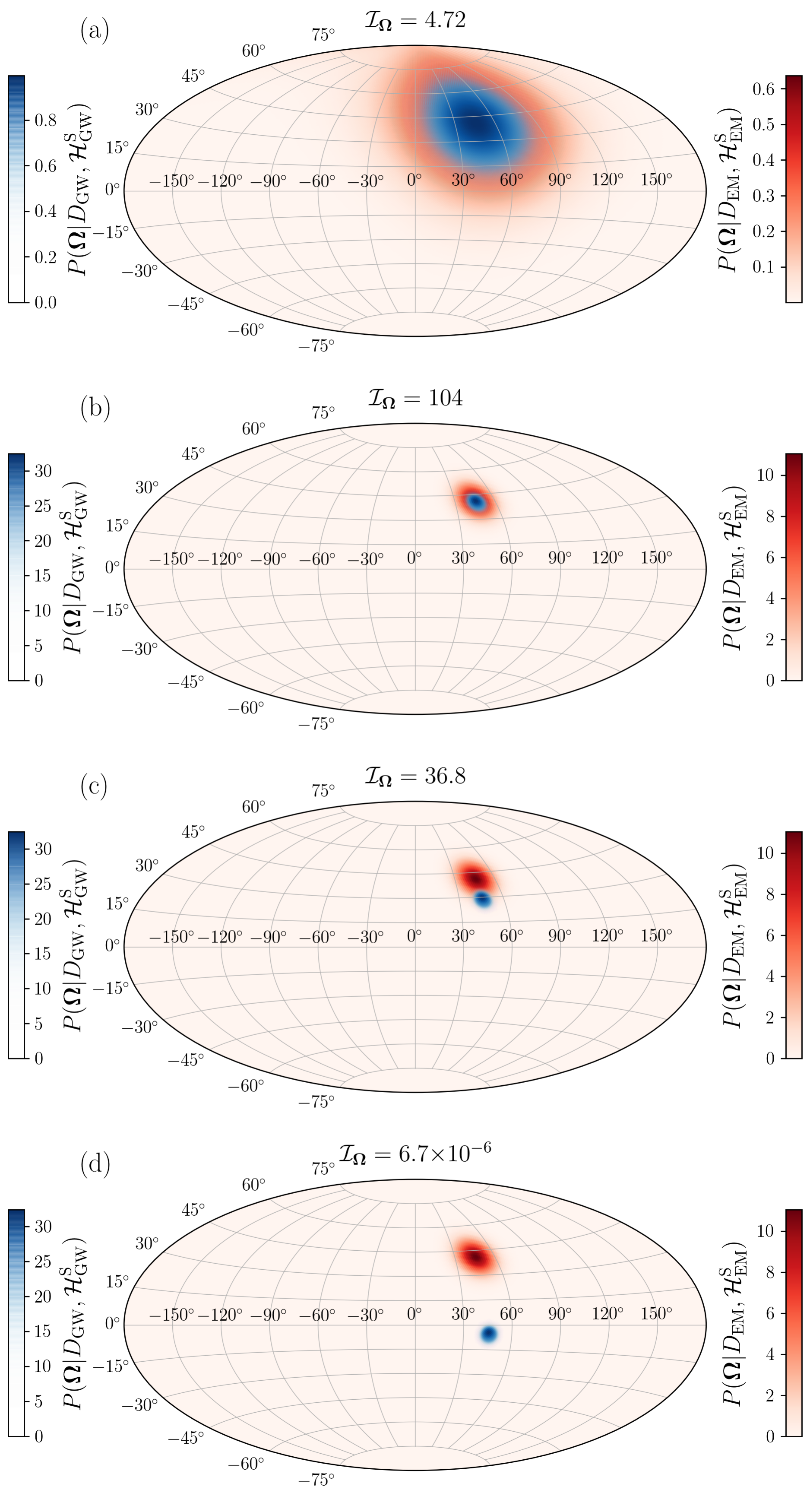}
\caption{Examples of the spatial overlap $\I_{\BOmega}$. A
blue (red) density map shows the probability per pixel for the GW (EM)
detection. $\I_{\BOmega}$ is calculated by numerical integration over these
pixels with an all-sky uniform prior. The posteriors are computed
over an array of pixels, each with equal area, using the HEALPix
projection \citep{gorski2005healpix}.}
\label{fig_spatial_examples}
\end{figure}

To illustrate the subtleties of $\I_{\BOmega}$ and provide some intuition, in
Fig.~\ref{fig_spatial_examples}, we show four examples varying the size of the
uncertainty region and angular separation of the means of the EM and GW sky
localizations. For all examples, a uniform all-sky prior is used. In
Fig.~\ref{fig_spatial_examples}(a), the means of both posteriors are aligned,
but the uncertainty on both is large with respect to the all-sky prior;
therefore, $\I_{\BOmega}$ is greater than one, but not large enough to be of
note. For Fig.~\ref{fig_spatial_examples}(b), $\I_{\BOmega}$ strongly indicates
the two detections are from the same event: the means are aligned and the
uncertainties are small with respect to the all-sky prior. In
Fig.~\ref{fig_spatial_examples}(c) and (d), the means of the distributions are
not aligned. While in (c) this results in modest evidence in favor of a common
event, the separation is sufficiently wide in (d) to strongly disfavor a
common source.

To help guide our intuition, we can also calculate $\I_{\BOmega}$
for the simplified case where the posterior distributions on the sky are
uniform distributions, i.e.\ constant inside the sets $\Pi\GW$ and $\Pi\EM$ and
zero outside. Labeling $\Delta(\Pi)$ the area of set $\Pi$ in square radians,
we obtain
\begin{align}
\begin{split}
\I_{\BOmega}(D\GW, D\EM)
&=4\pi\frac{\Delta(\Pi\GW \cap \Pi\EM)}{
\Delta(\Pi\GW) \Delta(\Pi\EM)}\,.
\end{split}
\label{eqn_sky_approx}
\end{align}
The $4\pi$ prefactor comes from the all-sky prior and acts as a metric to
compare the size of the overlap. For example, if $\Pi\GW$ is entirely contained
within $\Pi\EM$, then $\I_{\BOmega}(D\GW, D\EM) = 4\pi / \Delta(\Pi\EM)$: the
Bayes factor is determined entirely by the fraction of the sky covered by the
uncertainty on the EM detections (or vice versa if the EM posterior is
contained within the GW posterior).

\newpage
\subsubsection{The spatial and temporal odds}
\label{sec_final_odds}

To calculate the odds in this example via Eq.~\eqref{eqn_odds}, we require the
prior odds.  We consider a Poisson point process which produces events
detectable via either (or both) their GW or EM emission with a total rate $R$
per unit time, acting during the co-observing time $T$.  Further, we let
$R=R\GW+R\EM+R\GWEM$: the total rate is the sum of the rates of events
detectable only in GW, events detectable only in EM, and events jointly
detectable in both.  Then $\Hc$ refers to a signal seen by both detectors, and
$\Hss$ to signals detected in one or other, but not both.  Choosing $T$ such
that $RT\ll1$, and defining $\mathrm{Poisson}(1; \lambda)$ to be the
probability of one event given an expected number of signals $\lambda$, we
obtain
\begin{align}
\begin{split}
\frac{P(\Hc)}{P(\Hss)} &=
\frac{\mathrm{Poisson}(1; R\GWEM T)}
{\mathrm{Poisson}(1; R\GW T)\mathrm{Poisson}(1; R\EM T)}\\
 &\approx\frac{R\GWEM}{R\GW R\EM T}\,.
\end{split}
\end{align}
This prior odds clearly depends on the co-observing time $T$.  Combining this
with the spatial and temporal Bayes factor (Eq.~\eqref{eqn_spatial} and
\eqref{eqn_time_approximation_full}) then gives
\begin{align}
\Odds_{\Common/\SignalSignal}(D\subA, D\subB) =
\frac{R\GWEM}{R\GW R\EM}
\frac{1}{\intDeltat}
\I_{\BOmega}\,,
\label{eqn_odds_approximation}
\end{align}
which is not dependent on the co-observing time.  One special case is when
$R\GW \simeq R\GWEM \ll R\EM$, i.e.\ if signals detectable in EM only are much
more frequent than in GW, but we otherwise have little information on the rates of
GW detections with or without EM counterparts. This may typically occur if our
estimates of $R\GW$ and $R\GWEM$ are based on $\mathcal{O}(1)$ detection. The
odds are then proportional to $1/(R\EM\intDeltat)$, which reproduces the
temporal Bayes factor Eq.~\eqref{eqn_time_approximation_full} setting $T =
1/R\EM$, i.e.\ the waiting time between EM detections (where the great majority
have no GW counterpart).

As can be expected intuitively, the association becomes less significant if the
$\Delta t$ prior is broader or the prior background rate of signals is
higher, but increases with the prior expected rate of joint detections.

\subsection{Application to GW170817 and GRB~170817A}
\label{sec_gw170817_grb170817a}

We now apply the example calculated in Sec.~\ref{sec_example} to GW170817 and
GRB~170817A, the result of which can be compared with \citet{gwgbm}. We note that,
the calculation presented here could be improved by using the full joint
distribution without making assumptions that allow the result to be
factorized, and including other pertinent model parameters such as the
luminosity distance (for \emph{Fermi}-GBM, this may be as simple as estimating
the range of conceivable values).

The sky localization for the BNS inspiral and short GRB can be seen in Fig.~1
of \citet{gwgbm}. Using the published localization \texttt{FITS} files
\citep{gw170817localization, goldstein2017} and a uniform prior distribution on
the whole sky, Eq.~\eqref{eqn_spatial} yields $\I_{\BOmega}=32.4$.  The
spatial overlap alone provides moderate support for the common-event model, the
main limitation being the uncertainty on the localization of GRB~170817A.

If we did not have the actual \texttt{FITS} files, we could still use the
published 90\% confidence intervals for the sky localization of GW170817 and
GRB~170817A and take the localization posterior distributions to be uniform
within those intervals. The intervals cover respectively
$\SI{28}{\textrm{deg}\squared}$ \citep{multimessenger} and
$\SI{1100}{\textrm{deg}\squared}$ \citep{goldstein2017} and the GW170817
interval is entirely contained within GRB~170817A. A straightforward
application of Eq.~\eqref{eqn_sky_approx} then yields an approximate spatial
Bayes factor of $\I_{\BOmega} = 37.5$, which is close to the exact value. Since
we do have the full posteriors, however, we can repeat this calculation with
different confidence levels.  We find that $\I_{\BOmega}$ can be biased by a
factor of a few in both directions, depending on what confidence level is used;
the 90\% interval just happens to produce a particularly close number.

In calculating the odds from Eq.~\eqref{eqn_odds_approximation}, there are
large uncertainties on the three rates. However, the rate of short GRB
detections by \emph{Fermi}-GBM is well known and must satisfy $R\Fermi \approx
R\EM + R\GWEM$. One could model the uncertainties to produce beaming and volume
corrected estimates for these rates (see e.g.\ \citet{fong2015, siellez2016}).
However, for a simple estimate we assume that $R\GWEM$ and $R\GW$ are of
similar magnitude, and take $R\EM$ to be approximately $R\Fermi=0.124$~per-day
\citep{gwgbm}.  Then using $[\Deltatmin, \Deltatmax]= [-1, 5]$~s (as used in
the \citet{abadie2012} search) in Eq.~\eqref{eqn_odds_approximation}, the odds,
including the spatial overlap, are $\Odds_{\Common/\SignalSignal}(D\GW, D\EM)
\gtrsim 10^{6}$: the odds provide decisive evidence that the two detections
originate from the same event.

These numbers are consistent with the \emph{p}-values estimated in
\citep{gwgbm}: the time overlap dominates, the spatial part is small but
supports the hypothesis, and the overall factor is highly significant (the
total \emph{p}-value was found to be $5\times10^{-8}$ \citep{gwgbm}).

\subsection{Comparison with p-values}
\label{sec_comparison}

There are parallels that can be drawn between the odds calculated in
Sec.~\ref{sec_example} and the \emph{p}-value approach of \citep{gwgbm}; namely,
the spatial overlap Eq.~\eqref{eqn_spatial} with the $\mathcal{S}$ statistic
and the form of the temporal overlap (i.e.\ inversely proportional to the
background rate).

However, the two methods are not equivalent and the numerical values themselves
cannot be directly compared, as they answer different questions.  The odds is
exactly our relative degrees of belief for the common- vs.\ distinct-source
hypotheses, given the assumptions made in the calculation; the
\emph{p}-value tests whether the data is consistent or not with the null
(distinct-source) hypothesis, and is typically interpreted as the rate at which
a rule for deciding the significance of a joint detection leads to false
positives \citep{finn1999}.

Finally, a more practical difference is that while Eq.~\eqref{eqn_spatial} can
be directly interpreted as a Bayes factor for the spatial overlap, interpreting
the $\mathcal{S}$ statistic requires numerical calculation of the background by
randomly rotating a set of observed short GRB sky localizations.

\section{Conclusions}

We introduce a Bayesian model comparison approach to estimating
our confidence that two multimessenger observations are due to a common source
as opposed to an accidental coincidence of distinct sources.  The primary
result of this work is Eq.~\eqref{eqn_third_full}, which generically allows the
calculation of the Bayes factor (and, hence, the odds) from the joint posterior
distributions of common model parameters inferred independently from two data
sets. This approach forces us to recognize the conditions under which the
contributions to the Bayes factor can be factorized.

We provide an example where the spatial and temporal overlap calculation can be
approximately factorized for two independent observations with isotropic
observatories and apply the result to GW170817 and GRB~170817A. We find decisive
evidence in favor of their association, consistent with \citet{gwgbm}.

\acknowledgements

The authors are grateful to Michael Briggs, Collin Capano, Sebastian Khan,
Badri Krishnan, Francesco Pannarale, Yafet Sanchez Sanchez, Karelle Siellez, Grant Meadors and members
of the LIGO and Virgo collaborations for useful comments during the preparation
of this work. EB and TDC are supported by an appointment to the NASA
Postdoctoral Program at the Goddard Space Flight Center, administered by
Universities Space Research Association under contract with NASA.

\bibliographystyle{yahapj}
\bibliography{bibliography}

\begin{thebibliography}{}
\providecommand\natexlab[1]{#1}
\providecommand\JournalTitle[1]{#1}

\bibitem[{Aartsen {et~al.}(2014)Aartsen, Ackermann, Adams,
  {et~al.}}]{aarsten2014}
Aartsen, M.~G., Ackermann, M., Adams, J., {et~al.} 2014,
  \href{http://dx.doi.org/10.1103/PhysRevD.90.102002}{\JournalTitle{Phys. Rev.
  D}, 90, 102002}

\bibitem[{{Abadie} {et~al.}(2012){Abadie}, {Abbott}, {Abbott},
  {et~al.}}]{abadie2012}
{Abadie}, J., {Abbott}, B.~P., {Abbott}, R., {et~al.} 2012,
  \href{http://dx.doi.org/10.1088/0004-637X/760/1/12}{\JournalTitle{\apj}, 760,
  12}

\bibitem[{{Abbott} {et~al.}(2017{\natexlab{a}}){Abbott}, {Abbott}, {Abbott},
  {et~al.}}]{gwgbm}
{Abbott}, B.~P., {Abbott}, R., {Abbott}, T.~D., {et~al.} 2017{\natexlab{a}},
  \href{http://stacks.iop.org/2041-8205/848/i=2/a=L13}{\JournalTitle{\apjl},
  848, L13}

\bibitem[{{Abbott} {et~al.}(2017{\natexlab{b}}){Abbott}, {Abbott}, {Abbott},
  {et~al.}}]{gw170817}
---. 2017{\natexlab{b}},
  \href{http://dx.doi.org/10.1103/PhysRevLett.119.161101}{\JournalTitle{\prl},
  119, 161101}

\bibitem[{{Abbott} {et~al.}(2017{\natexlab{c}}){Abbott}, {Abbott}, {Abbott},
  {et~al.}}]{multimessenger}
---. 2017{\natexlab{c}},
  \href{http://stacks.iop.org/2041-8205/848/i=2/a=L12}{\JournalTitle{\apjl},
  848, L12}

\bibitem[{{Baret} {et~al.}(2012){Baret}, {Bartos}, {Bouhou},
  {Chassande-Mottin}, {Corsi}, {Di Palma}, {Donzaud}, {Drago}, {Finley},
  {Jones}, {Klimenko}, {Kouchner}, {M{\'a}rka}, {M{\'a}rka}, {Moscoso}, {Papa},
  {Pradier}, {Prodi}, {Raffai}, {Re}, {Rollins}, {Salemi}, {Sutton}, {Tse},
  {Van Elewyck}, \& {Vedovato}}]{baret2012}
{Baret}, B., {Bartos}, I., {Bouhou}, B., {et~al.} 2012,
  \href{http://dx.doi.org/10.1103/PhysRevD.85.103004}{\JournalTitle{\prd}, 85,
  103004}

\bibitem[{Coulter {et~al.}(2017)Coulter, Foley, Kilpatrick,
  {et~al.}}]{coulter2017}
Coulter, D.~A., Foley, R.~J., Kilpatrick, C.~D., {et~al.} 2017,
  \href{http://dx.doi.org/10.1126/science.aap9811}{\JournalTitle{Science}}

\bibitem[{{Finn}(1998)}]{finn1998}
{Finn}, L.~S. 1998, in Second Edoardo Amaldi Conference on Gravitational Wave
  Experiments, ed. E.~{Coccia}, G.~{Veneziano}, \& G.~{Pizzella}, 180

\bibitem[{Finn {et~al.}(1999)Finn, Mohanty, \& Romano}]{finn1999}
Finn, L.~S., Mohanty, S.~D., \& Romano, J.~D. 1999,
  \href{http://dx.doi.org/10.1103/PhysRevD.60.121101}{\JournalTitle{\prd}, 60,
  121101}

\bibitem[{{Fong} {et~al.}(2015){Fong}, {Berger}, {Margutti}, \&
  {Zauderer}}]{fong2015}
{Fong}, W., {Berger}, E., {Margutti}, R., \& {Zauderer}, B.~A. 2015,
  \href{http://dx.doi.org/10.1088/0004-637X/815/2/102}{\JournalTitle{\apj},
  815, 102}

\bibitem[{Gelman {et~al.}(2013)Gelman, Carlin, Stern,
  {et~al.}}]{gelman2013bayesian}
Gelman, A., Carlin, J.~B., Stern, H.~S., {et~al.} 2013, {Bayesian Data
  Analysis} (CRC press)

\bibitem[{Goldstein {et~al.}(2017)Goldstein, Veres, Burns,
  {et~al.}}]{goldstein2017}
Goldstein, A., Veres, P., Burns, E., {et~al.} 2017,
  \href{http://dx.doi.org/10.3847/2041-8213/aa8f41}{\JournalTitle{\apjl}, 848,
  L14}

\bibitem[{Gorski {et~al.}(2005)Gorski, Hivon, Banday,
  {et~al.}}]{gorski2005healpix}
Gorski, K.~M., Hivon, E., Banday, A.~J., {et~al.} 2005, \JournalTitle{\apj},
  622, 759

\bibitem[{Gregory(2005)}]{gregory2005bayesian}
Gregory, P. 2005, {Bayesian Logical Data Analysis for the Physical Sciences: A
  Comparative Approach with Mathematica{\textregistered} Support} (Cambridge
  University Press)

\bibitem[{Haris {et~al.}(2017)Haris, Mehta, Kumar, \& Ajith}]{haris2017}
Haris, K., Mehta, A.~K., Kumar, S., \& Ajith, P. 2017, private communication

\bibitem[{{Keivani} {et~al.}(2015){Keivani}, {Fox}, {Te{\v s}i{\'c}}, {Cowen},
  \& {Fixelle}}]{amon2015}
{Keivani}, A., {Fox}, D.~B., {Te{\v s}i{\'c}}, G., {Cowen}, D.~F., \&
  {Fixelle}, J. 2015, \JournalTitle{ArXiv e-prints},
  \href{http://arxiv.org/abs/1508.01315}{{\sffamily arXiv:1508.01315
  [astro-ph.HE]}}

\bibitem[{Kelley {et~al.}(2013)Kelley, Mandel, \& Ramirez-Ruiz}]{kelley2013}
Kelley, L.~Z., Mandel, I., \& Ramirez-Ruiz, E. 2013,
  \href{http://dx.doi.org/10.1103/PhysRevD.87.123004}{\JournalTitle{Phys. Rev.
  D}, 87, 123004}

\bibitem[{Margutti {et~al.}(2017)Margutti, Berger, Fong,
  {et~al.}}]{margutti2017}
Margutti, R., Berger, E., Fong, W., {et~al.} 2017,
  \href{http://stacks.iop.org/2041-8205/848/i=2/a=L20}{\JournalTitle{\apjl},
  848, L20}

\bibitem[{Savchenko {et~al.}(2017)Savchenko, Ferrigno, Kuulkers,
  {et~al.}}]{savchenko2017}
Savchenko, V., Ferrigno, C., Kuulkers, E., {et~al.} 2017,
  \href{http://stacks.iop.org/2041-8205/848/i=2/a=L15}{\JournalTitle{\apjl},
  848, L15}

\bibitem[{{Siellez} {et~al.}(2016){Siellez}, {Boer}, {Gendre}, \&
  {Regimbau}}]{siellez2016}
{Siellez}, K., {Boer}, M., {Gendre}, B., \& {Regimbau}, T. 2016,
  \JournalTitle{ArXiv e-prints},
  \href{http://arxiv.org/abs/1606.03043}{{\sffamily arXiv:1606.03043
  [astro-ph.HE]}}

\bibitem[{Singer(2017)}]{gw170817localization}
Singer, L. 2017, {GW170817 sky localization},
  \url{https://dcc.ligo.org/LIGO-G1701985/public}

\bibitem[{Soares-Santos {et~al.}(2017)Soares-Santos, Holz, Annis,
  {et~al.}}]{soares2017}
Soares-Santos, M., Holz, D.~E., Annis, J., {et~al.} 2017,
  \href{http://stacks.iop.org/2041-8205/848/i=2/a=L16}{\JournalTitle{\apjl},
  848, L16}

\bibitem[{{Troja} {et~al.}(2017){Troja}, {Piro}, {van Eerten},
  {et~al.}}]{troja2017}
{Troja}, E., {Piro}, L., {van Eerten}, H., {et~al.} 2017,
  \href{http://dx.doi.org/10.1038/nature24290}{\JournalTitle{\nat}}

\bibitem[{{Urban}(2016)}]{urban2016monsters}
{Urban}, A.~L. 2016, {Ph.D.} dissertation, {University of Wisconsin-Milwaukee},
  \url{https://search.proquest.com/docview/1806791084}

\end{thebibliography}

\end{document}